\documentclass[letterpaper, 10 pt, conference]{ieeeconf}  

\IEEEoverridecommandlockouts                              
\makeatletter

\newcommand{\Rmnum}[1]{\expandafter\@slowromancap\romannumeral #1@}
\makeatother

\usepackage{graphics} 

\title{\LARGE \bf
The Child Prodigies of IEEE 802.11 Family
}

\author{Rohit Gandikota$^{1}$
\thanks{*This work is a summary of 4 papers in total on IEEE 802.11 Family. It is intended for non-profitable educational purpose only.}
\thanks{$^{1}$Rohit Gandikota is an Undergrad student in the Department of Avionics, Indian Institute of Space science and Technology, Kerala, India
        {\tt\small grohit0 at gmail.com}}%
 }
\begin{document}

\maketitle
\thispagestyle{empty}
\pagestyle{empty}

\begin{abstract}
WiFi was designed just to replace the wired connections, but it has changed the Internet as a whole. WiFi users are increasing day by day and the traffic from wireless devices is dominating the Internet. With increasing number of users and their expectations on QoS(Quality of Service) and data rates, there is a need for better WiFi techniques which can support the above mentioned challenges. The IEEE LAN/MAN standards committee (LMSC) has launched new protocols like IEEE 802.11ac, 802.11ad and 802.11ax. Understanding them is very important to use them efficiently and effectively to fulfill our expectations. In this paper we summarize few of the works \cite{c1}, \cite{c2}, \cite{c3} and \cite{c4} which talk about the above the mentioned protocol standards of WiFi that brought a significant improvements in WiFi's performance. Moreover we discuss the differences in the protocols and their applications. We also try to compare these three protocols with respect to their supported data rates, range of transmission and Robustness.   
\end{abstract}

\section{INTRODUCTION}
Wireless Networking has become a fundamental technology as important as computing itself. WiFi has been pushing it's limits on performance and user experience guaranteeing that it is keeping pace with the ever increasing demand of higher speeds. According to  \cite{c3} and \cite{c4}, WiFi is now capable of Multi-Gigabit data rates. \par With the advent of Bandwidth hungry applications like online gaming, HD video streaming and raw uncompressed image uploads, there is a need for more bandwidth. Two emerged standards which shook the wireless world: IEEE 802.11ac \cite{c5} and IEEE802.11ad \cite{c6} had targeted data rates faster than Gigabits for the first time. The history of initiation and development of these standards are presented in \cite{c7}.
\par In this paper we talk about 
\begin{itemize}
\item The emerged IEEE wireless protocols 802.11ac, 802.11ad and 802.11ax.
\item Key physical(PHY) layer features of the three protocols.
\item The need for Multi-Gigabit WiFi and how 802.11ac and 802.11ad has achieved it.
\item The three main challenges with telecommunication technologies and how 802.11ax has overcome them.
\item Other challenges and application cases to use these protocols.
\end{itemize}
The rest of the paper is organized as follows. In section \Rmnum{2} previous work on IEEE WLAN standards are discussed. In section \Rmnum{3} three protocols that revolutionized the WiFi, 802.11ac,802.11ad and 802.11ax are summarized in detail. In section \Rmnum{4} challenges associated with these protocols and the performances are compared and discussed. Finally in section \Rmnum{5} we conclude our discussion.    
\section{Previous Work}
The 802.11 designation refers to the IEEE’s WLAN standard, commonly called by it's trade name, WiFi.
The suffix indicates one alternative of the standard.The 802.11 standard defines both a Physical layer (PHY) and Media Access Control (MAC)
layer in the networking scheme. While the MAC tends to remain mostly the same, the PHY changes to
include the most recent wireless technology for greater speed and link reliability.\par  Commercial WiFi was born when the 11b version arrived in 1999, it facilitated the first widespread implementation of WLAN technology. The 802.11b standard is often considered the first generation, 802.11a is the second generation, 802.11g is the third generation, and 802.11n is the fourth generation. 802.11ac/ad will represent the fifth generation. IEEE 802.11n has been the most widely implemented variant found in WiFi access points, hotspots, and routers. The later generation of the standard, 802.11ac, 802.11ad 
defined a faster and more reliable technology. These generations are discussed below
\subsection{802.11b}
The first generation, 11b, used Direct Sequence Spread Spectrum (DSSS) to achieve data rates up to 11 Mbps in
a 20-MHz channel on the 2.4-GHz band. When the 802.11b standard was introduced in 1999, the speed of wireless at 11 Mbps was quite useful and faster than almost any other similar technology. But with the growth of the Internet, that speed soon became insufficient. A faster version was sought, bringing about major changes in the Physical (PHY) layer. 
\subsection{802.11a}
Second-generation 802.11a came after 1999. It was the first to use the 5-GHz ISM band and Orthogonal Frequency Division Multiplexing (OFDM) with 64 sub-carriers spaced at 312.5 kHz. Channel bandwidth was 20 MHz. Modulation types like Binary Phase Shift Keying (BPSK), Quadrature Phase Shift Keying (QPSK), 16-phase Quadrature Amplitude Modulation (16-QAM), and 64-phase Quadrature Amplitude Modulation (64-QAM) were defined. This permitted the data rate to increase to a maximum of 54 Mbps. While the 802.11a version was more robust because of the OFDM characteristics that reduced multipath reflections and the 5-GHz band led to less interference, the higher frequency still limited the range. This version was never popular despite its advantages.
\subsection{802.11g}
The 802.11g standard was approved in 2003. Technically, 11g is the same as 11a but operating in the 2.4 GHz band. Using the exact same OFDM and modulation concepts, it too has the capability to deliver up to 54 Mbps rates. This got quite popular than it's parent protocol 802.11a.
\subsection{802.11n}
The 11n standard is a further improvement over 11a/g. It has enhanced the PHY by adopting two major techniques
\begin{enumerate}
\item 40-MHz channel(Channel Bonding of two 20MHz channels)
\item Multiple-Input Multiple-Output (MIMO)  
\end{enumerate}

 These methods allow data rates to reach till 600 Mbps(10 times higher than 802.11g). MIMO is a technique of using multiple antennas as receivers and transmitters to achieve something called Spatial Division Multiplexing (SDM). SDM helps transmitting multiple data streams concurrently within the same 20 MHz or 40 MHz channels. This results in the data rates multiplied by a factor roughly equal to the number of simultaneous data streams. The 11n standard limits up to four transmit and four receive channels (4x4), although 1x2, 2x2, and 3x3 versions are more widely used. The 600 Mbps data rate is achieved using 4x4 MIMO with 64QAM modulation scheme in a 40-MHz channel.

\section{New Generation of IEEE 802.11}
 
As seen in the previous section, 802.11n has set a milestone in the WLAN protocols, taking the inspiration from this, two protocols 802.11ac and 11ad had emerged. Later in year 2016, LMSC(LAN / MAN Standards Committee) has established a Task Group which has released the first draft of a High Efficiency WLAN, IEEE 802.11ax. In this section, we discuss these three protocols.
\subsection{\textbf{802.11ac}}
IEEE 802.11ac Task Group has developed an amendment to 802.11 PHY and MAC layers to meet the growing demands of new wireless applications that require higher data rates. The two main features implanted into 802.11ac are
\begin{itemize}
\item 80/160 MHz Static and Dynamic Channel Bonding technique that bonds two 40/80 MHz channels. 
\item Multi-user Multi-Input Multi-Output (MU-MIMO)
\end{itemize}
According to \cite{c1}, dynamic channel bonding outperforms static method by 85\% when secondary channel are having moderate traffic load from the legacy(802.11 a/g/n) stations(STA). They also talk about the effects of secondary channel(CCA) sensitivity and primary channel positioning effects on 802.11ac's throughput.\par 
On the other hand \cite{c2} additionally talks about the MU-MIMO, enhanced RTS-CTS Mechanisms, Transmit beamforming method and Coding techniques of 802.11ac PHY and MAC. We will summarize these two papers in this subsection.
\par 802.11ac has tried to extend the static 40MHz and Dynamic 20/40 MHz bandwidth channel access schemes of 802.11n. As the channel bandwidth gets wider and as environment gets denser due to increase in wireless devices, it gets harder to find a clean 80 MHz wide channel. Also, since the internet backbone and all the stations use the older/legacy versions of WiFi, it is more likely for 802.11ac to share the channels with the legacy 802.11a/n devices operating in 20 or 40 MHz. As more and more 802.11a/n devices are operating in the secondary channels of 11ac, the performance of 11ac devices degrades. Therefore it is vital to use the frequency resources efficiently by adopting Dynamic channel bonding. Also the MIMO has a multiplied increase in the throughput. These are discussed below \par
\textbf{Static channel access:} Here the station reattempts to acquire the full clean 80MHz channel after a certain time according to the contention window's state. This happens when the secondary channels are busy due to the legacy stations in the same channel. Obviously this reduces the throughput.\par 
\textbf{Dynamic Channel Access:} Here the station may transmit data over a narrower channel of 20 or 40 MHZ depending on the CCA(Clear Channel Assessment) results of secondary results. This approaches utilizes the frequency efficiently. \cite{c1} shows simulations to prove that this method is 85\% better than the previous.\par
\textbf{MIMO:}  With multiple antennas we can achieve two things, Single User MIMO(SU-MIMO) and Multi User MIMO(MU-MIMO). SU-MIMO is a concept where a device having 802.11ac is connected to a device with multiple receivers. Here we can use spatial multiplexing. Only a single user at a time. While in MU-MIMO there are multiple devices connected at a single time. In 802.11ac they have introduced only Downlink MU-MIMO(DL-MU-MIMO). To achieve this we need some sort of feedback info about the user's Channel State Information(CSI). This is achieved by transmit beamforming.\par
\textbf{Transmit Beamforming:} This helps in a directed antenna's higher gains compared to an omni directional antenna. This is done by process called \textit{Sounding}. The Access Point (AP) sends training info to the users and wait for the info feedback from them. This feedback is then used to calculate some sort of weight/steering matrix within the AP. This matrix is used to precode the transmissions by creating a set of steered beams to optimize reception at one or multiple users.\par
\textbf{Modulation and Coding Schemes:} The PHY data subcarriers in 802.11ac are modulated using Binary Phase Shift Keying (BPSK), Quadrature Phase Shift Keying (QPSK), 16-Quadrature Amplitude Modulation (QAM), 64-QAM, and 256-QAM. Note that 256-QAM is
just supported by 802.11ac.Coding schemes like Forward Error Correction (FEC) coding is used with coding rates of 1/2, 2/3, 3/4, and 5/6. Use of Binary Convolutional Coding(BCC) is mandatory, but Low-Density Parity-Check Coding (LDPC) is optional.

\newpage
\subsection{\textbf{802.11ad}}
IEEE 802.11ad \cite{c3} is a new amendment to the IEEE WLAN protocol family, built for enhancements in gigabit throughput in 60GHz bands. In this band typically 7GHz of spectrum is available compared to 83.5 MHz spectrum in 2.4 GHz band. This standard defines 4 channels, each with 2.16 GHz spevtrum bandwidth, for operation at 60 GHz band. These are 54 times wider than the 40 MHz bonded channels available in 802.11n.\par 
This standard has OFDM which allows SQPSK, QPSK, 16-QAM, and 64-QAM modulation(similar modulation schemes as 802.11ac) with the maximum achievable data rate of 6.756 Gbps. SC PHY is low on power consumption and focuses on small form factor devices like handsets. SC uses \(\pi\)/2-BPSK, \(\pi\)/2-QPSK, and \(\pi\)/2-16-QAM modulation with the maximum achievable data rate of 4.620 Gbps. The coding techniques are similar to 802.11ac.\par 
Here in 802.11ad multi-gigabit is achieved by using a large chunk of spectrum with simple modulation schemes(BPSK, QPSK), while in 802.11ac it is achieved by exploiting the concept of sending more bits per symbol(256-QAM) and use of SDM since the bandwidth is limited(160Hz maximum with channel bonding).\par 
The 802.11n data rates range from 6.5–600 Mbps, achieved through various combinations of  code rate, modulation scheme, guard interval, channel bandwidth and number of spatial streams. 802.11ac PHY data rates range from 6.5 Mbps to 6.93 Gbps. Whereas 802.11ad data rates range from 385 Mbps to 6.7 Gbps, achieved through combinations
of code rate and modulation scheme. For the BPSK modulation scheme, PHY data rate
achieved by 802.11ad using 2.16 GHz channel bandwidth and 802.11n employing 20 MHz
channel bandwidth is 385 Mbps and 6.5 Mbps, respectively, i.e., a 58 times increase in the PHY data rate, leveraging larger channel bandwidth.\cite{c3}\par  
802.11ad uses similar \textbf{RTS/CTS mechanism}(Request to Send/ Clear to Send) as used in 802.11ac. If an 802.11ac/ad AP is nearby other legacy devices, it is possible that the 20 MHz primary channel of any of the latter ones is anywhere within 80/160/2160 MHz of the 11ac/ad AP. This means that the different APs and their clients can transmit at overlapping times on different sub-channels, thus leading to collisions or deferrals.\par 
To overcome this problem, 802.11ad defines a new upgraded handshake to properly handle both static and dynamic channel allocation. This handshake(modified RTS/CTS) consists of a  mechanism that gives information about the current amount of available bandwidth. \cite{c2} show how the modified RTS/CTS protocol works with the following example : Consider a scenario in which an initiating AP(802.11ac) wants to transmit data to a client through an 80 MHz channel. The AP first checks if the channel is idle or not. If it is idle, AP transmits multiple RTS in the 802.11a PPDU format (one RTS for each
20 MHz channel). Therefore, it is expected that every nearby device (legacy or 802.11ac) can receive an RTS on its primary channel. Each of these devices then sets their NAV. Before a client replies with a CTS, it checks if any of the channels
in the 80 MHz band is busy. The client only replies with a CTS on those channels
that are idle, and reports the total bandwidth of the duplicated CTS. As with the RTS, the CTS is sent in an 802.11a PPDU format and is duplicated across the different 20 MHz channels that are idle.
\newline
\subsection{\textbf{802.11ax}}
To tackle problems like exponential growth in traffic, increase in the expectations of users and rapid growth of wireless devices being established, LMSC launched a High Efficiency WLAN Task Group AX. By 2019 this group aims to develop IEEE 802.11ax. This standard increases transmission rates along with improving efficiency of channel usage in case of dense deployment.\par 
While 802.11ac offers only Downlink MU-MIMO, 11.ax offers both Uplink and Downlink MU MIMO. It brings a new Orthogonal Frequency-Division Multiple Access(OFDMA) to WiFi. Thus 11ax aims to increase the throughput by 4 times. But for this it has to keep few parameters on check like drastic increase in the energy consumption. Since all the devices are battery powered, power efficiency also becomes one of the parameters.\par This task group mainly aim to solve 3 peculiarities.

\begin{figure}

\includegraphics{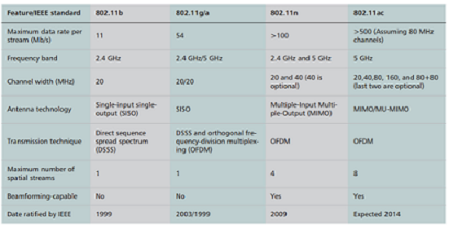}
\caption[Sèries temporals de vent i direccions]{Comparision between 802.11 family}
\label{Serie}

\end{figure}
\begin{itemize}
\item Dense WiFi deployment like, stadiums and public areas.
\item Related Traffic due to the above like, real-timing streams and file download.
\item Huge Upload of numerous photos, videos and documents to social networks.
\end{itemize}
802.11ax is set to bring some modifications in the PHY. It introduces OFDMA which is OFDM with longer symbols. This increases robustness and granularity. 802.11ax also Guard Interval between the symbols to decrease inter symbol interference. But this increases overhead. To improve the situation, OFDMA let's sender use only best set of tones, so called resource units(RU) through a feedback. According to \cite{c8}, this feedback gives 50\% throughput gain in indoor and outdoor scenarios with respect to random allocation of channels.
\par There is also a change in the frame format. Similar to 11n and 11ac, frame starts with preamble. This preamble has two parts, High Efficiency one and the Legacy one. The legacy preamble is easily decoded by the legacy devices while the HE preamble is decoded only by the 11ax devices. To distinguish between HE and Legacy preamble, HE preamble starts with repetition of LSIG. As shown in Fig.2.

\begin{figure}

\includegraphics{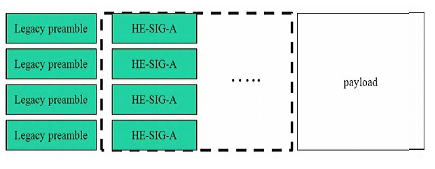}
\caption[Sèries temporals de vent i direccions]{Legacy Preamble and HE-SIG-A frame structure.}
\label{Serie}

\end{figure}
\par
Let us see modulation schemes and coding schemes introduced in 802.11ax. First let us consider 1024-QAM modulation. Since it needs higher SNR, it can benefited in indoor environment. Second 11ax introduces Dual sub-carrier modulation(DCM). This allows same transmissions in two channels, which are far apart in frequency wise. DCM improves transmission robustness in presence of sub-band interference.\par 
802.11ax also uses UL-DL-MIMO to increase the throughput in multiple amount. This concept is discussed in 11ad. In addition to this 11ax also provides Aloha-like random OFDMA UL-MU channel access.Although Aloha shows worse results, it makes the process easy.\par
To improve the performance of 802.11ax in case of dense scenarios, TGax enhances Overlapped BSS operation and spatial reuse. It also introduces multiple BSSID support, which allows sending identical information for all BSSs simultaneously via a common beacon.
\section{Challenges}
\subsection{Collision due to secondary CCA}
We know that the secondary channel has a lesser CCA sensitivity level compared to primary channel.The primary channel requires a receiver to be able to detect the structure of the signal such as the preamble of the signal. The secondary channel access can be achieved by a simple energy detection scheme and therefore extra decoding is not required. But this can cause collisions.\par  Let us consider an example from \cite{c1}, which assumes that the 802.11ac has the primary channel CCA sensitivity level of -82 dBm and the secondary channel CCA sensitivity level of -62 dBm. Therefore, if the received signal power at the secondary channel of the 802.11ac station is higher than -82 dBm but lower than -62 dBm, the 802.11ac station will consider the secondary channel to be idle and transmit an 80 MHz wide signal, which may collide with the transmission on the secondary channel.\par 
One way to address this problem is to have better CCA in the secondary channels so that the 802.11ac stations can better detect the signals in the secondary channels and avoid the
collisions. One way to achieve lower secondary channel sensitivity is to exploit the transmission signal structure such as the preamble structure or the OFDM signal structure. 

\subsection{Hardware Complexity}
This was addressed in \cite{c3}. 802.11ad systems require simpler hardware compared to 802.11ac, due to simpler modulation schemes and use of only one stream of data (SISO vs. MIMO). To have multiple independent data streams in 802.11ac, multiple RF and base-band chains are required. In practice, for better radio link performance, the number of RX and TX chains may be larger than the number of desired streams, NS (i.e., number of independent and separately encoded transmit signals
or streams). This implies more than NS times increase in power and RF chip/device area.
\subsection{Scenario usages}
This is considered as a challenge because not all standards are compatible for every scenario. Understanding this very important. Firstly considering 802.11ac, it seems to be more appropriate
for longer-range applications. As seen previously, 802.11ac requires –48 dBm receive sensitivity
with 256-QAM modulation and up to 8 spatial streams to achieve multi-gigabit WiFi. To deploy
the highest data rates of 802.11ac, AC-powered units are more suitable. Since the obstruction
loss at 5 GHz is lower than at 60 GHz(802.11ad), multigigabit 802.11ac is more appropriate for home scale (both line-of-sight and non-line-of-sight) wireless applications where portability is not a bottleneck. \par 
Talking about 802.11ad,  At 60 GHz spectrum, radio signals suffer from higher propagation and atmospheric loss compared to 5 GHz. Note that a general rule of thumb is every 6 dB increase in propagation loss halves the coverage distance. Furthermore, obstruction loss is significant at 60 GHz. Therefore, 802.11ad is more appropriate for line-of-sight, room-scale, low-cost, short-range,
very-high-throughput applications, such as in-room uncompressed and lightly compressed multimedia
wireless display, sync data/file transfer etc. \par 
Finally the new 802.11ax is more suitable for a dense network situation like, stadiums and public places like airport, shopping mall etc. They are more appropriate for dense deployment of WiFi. They seem to manage the traffic associated with these type scenarios.
\par Therefore we can not decide a particular standard to be better as the scenario we are using is also important. Along with our requirements, using a standard with more bandwidth we want is not really necessary.

\section{CONCLUSIONS}
We investigate the different standards of WLANs that have been announced by LMSC. To use a standard to it's full extent we need to consider the scenario and our requirements. We discussed their main contributions to the WiFi. We also talked about various challenges and how each standard overcome. WiFi is booming rapidly and is used for wireless connectivity between devices in a surfeit of scenarios. This is also causing a boom in the requirements and traffic. Hence it is interesting to see how future standards will be able to meet these.

\addtolength{\textheight}{-12cm}   




\end{document}